\documentclass[aps,prd,showpacs,nofootinbib,tightenlines]{revtex4-1}
\usepackage{amsmath}
\usepackage{amssymb}
\usepackage{epsfig}
\usepackage{graphicx}
\usepackage{color}
\usepackage{CJK,upgreek,fancyhdr}
\usepackage{hyperref}
\definecolor{Red}{rgb}{1.,0.,0.}

\definecolor{Blue}{rgb}{0.,0.,1.}

\definecolor{nicered}{rgb}{0.7,0.1,0.1}
\definecolor{nicegreen}{rgb}{0.1,0.5,0.1}

\hypersetup{colorlinks,citecolor=nicegreen,linkcolor=nicered}
\begin{document}
\begin{CJK*}{GB}{gbsn}
\title{$S$-wave resonance contributions to the $B^0_{(s)}\to \eta_c{(2S)}\pi^+\pi^-$
in the perturbative QCD factorization approach}
\author{Ai-Jun Ma (马爱军)$^1$}         \email{theoma@163.com}
\author{Ya Li (李亚)$^1$}               \email{liyakelly@163.com}
\author{Wen-Fei Wang (王文飞)$^2$}   \email{wfwang@sxu.edu.cn}
\author{Zhen-Jun Xiao (肖振军)$^{1,3}$} \email{xiaozhenjun@njnu.edu.cn}

\affiliation{$^1$ Department of Physics and Institute of Theoretical Physics,
                          Nanjing Normal University, Nanjing, Jiangsu 210023, P.R. China}
\affiliation{$^2$ Institute of Theoretical Physics, Shanxi University, Taiyuan, Shanxi 030006, P.R. China}
\affiliation{$^3$ Jiangsu Key Laboratory for Numerical Simulation of Large Scale Complex
Systems, Nanjing Normal University, Nanjing, Jiangsu 210023, P.R. China}
\date{\today}
\begin{abstract}
By employing the perturbative QCD (PQCD) factorization approach,
we study the quasi-two-body $B^0_{(s)}\to \eta_c{(2S)}\pi^+\pi^-$ decays,
where the pion pair comes from the $S$-wave resonance $f_0(X)$.
The Breit$-$Wigner formula for the $f_0(500)$ and $f_0(1500)$ resonances, and
the Flatt\'e model for the $f_0(980)$ resonance are adopted to parameterize the time-like
scalar form factors in the two-pion distribution amplitudes.
As a comparison, Bugg's model is also used for the wide $f_0(500)$ in this work.
For decay rates, we found the following PQCD predictions:
(a) $ {\cal B}(B^0_s\to \eta_c(2S) f_0(X)[\pi^+\pi^-]_s )=\left ( 2.67^{+1.78}_{-1.08} \right )\times 10^{-5}$
when the contributions from $f_0(980)$ and $f_0(1500)$ are all taken into account;
(b) ${\cal B}(B^0\to \eta_c(2S) f_0(500)[\pi^+\pi^-]_s)=
\left ( 1.40 ^{+0.92}_{-0.56} \right ) \times 10^{-6}$ in the Breit-Wigner model
and $ \left ( 1.53 ^{+0.97}_{-0.61} \right ) \times 10^{-6}$ in the Bugg's model.
\\
\\
Key words:  PQCD factorization approach, Two-pion distribution amplitudes, Quasi-two-body
decay
\end{abstract}

\pacs{13.25.Hw, 13.20.He, 13.30.Eg}

\maketitle


\section{Introduction}\label{sec:1}

The study of the three-body hadronic $B$ meson decays can help us understand the standard model and search for
the possible effects of new physics. Experimentally, quite a number of channels have been measured by collaborations like
BaBar~\cite{prl90-091801,prd70-092001,prd72-072003,prd80-112001,prd79-072006,prd78-052005},
Belle~\cite{prd75-012006,prl96-251803,prd71-092003,prd79-072004} and LHCb~
\cite{prd90-112004,prl112-011801,prl111-101801,prd86-052006,prd89-092006,prd92-032002,prd87-052001,npb-871403,prd90-012003,plb742-38,1702.08048}.
Theoretically, there are several approaches working in this field, for instance, the QCD factorization~
\cite{plb622-207,prd74-114009,prd79-094005, prd81-094033,appb42-2013,prd66-054015,prd72-094003,prd76-094006,prd88-114014,prd89-074025,prd94-094015,prd89-094007,npb899-247,epjc75-536,prd87-076007,1512.09284,prd70-034033},
 the perturbative QCD (PQCD) approach~\cite{plb561-258,prd70-054006,prd89-074031,prd91-094024,plb763-29,epjc76-675,1611.08786,prd95-056008,1701.02941},
and some methods based on symmetry principles~\cite{plb564-90,prd72-075013,prd72-094031,prd84-056002,
 prd84-034040,plb727-136,plb726-337,prd89-074043,plb728-579,ijmpa29-1450011,prd91-014029,prd89-094013,prd92-054010}.
 The aim of those studies is to understand the the resonant and nonresonant contributions, as well as
 the final state interactions (FSIs)~\cite{1512.09284,prd89-094013} in three-body $B$ decays.
But it is still in the early stage for both the theoretical studies and the experimental
measurements in studying those decays.

The PQCD factorization approach  is one of the major theoretical frameworks to deal
with the two-body hadronic $B$ meson decays~\cite{plb504-6,ppnp51-85}.
Very recently, some three-body hadronic $B$ meson decays have been studied by employing the
PQCD factorization approach, for example in Refs.~\cite{plb561-258,prd70-054006,prd89-074031,prd91-094024,plb763-29,epjc76-675,1701.02941,1611.08786,prd95-056008}.
For the cases of three-body decays, however, the previous PQCD approach~\cite{plb504-6,ppnp51-85}
should be  modified by introducing the two-meson distribution amplitudes~\cite{fp42-101,prl81-1782,npb555-231,sjnp38-289}
to describe the selected pair of final state mesons
due to the following reason discussed in~\cite{plb504-6,ppnp51-85}:
the contribution from the
direct evaluation of hard b-quark decay kernels containing two virtual gluons is generally
power suppressed, and the dominant contribution comes most possibly from the region where the
two energetic light mesons are almost collimating to each other with an invariant mass below
$O(\bar\Lambda m_B)$( $\bar\Lambda=m_B-m_b$, means the $B$ meson and $b$ quark mass difference).
Then, the typical PQCD factorization formula with the crucial nonperturbative
input of two-hadron distribution amplitudes for a $B\to h_1h_2h_3$ decay
amplitude can be written symbolically in the form of
\begin{eqnarray}
\mathcal{A}=\phi_B\otimes H\otimes \phi_{h_1h_2}\otimes\phi_{h_3}.
\end{eqnarray}
Here the hard kernel $H(x_i,b_i,t)$ contains the contributions from one hard gluon exchange diagrams only,
the nonperturbative inputs $\phi_B(x,b)$, $\phi_{h_1h_2}(z,\omega)$, $\phi_{h_3}(x_3,b_3)$ are the distribution
amplitudes for  the $B$ meson, the $h_1$-$h_2$ pair and the $h_3$ meson
respectively, while the symbols $\otimes$ mean the convolution integration over the
variables of the momentum fractions $(x,z,x_3)$  and the conjugate space coordinates $b_i$ of $k_{\rm iT}$.
With the help of the two-pion
distribution amplitudes, many works have been done for quasi-two-body decays, the parameters in the $S$-wave
and $P$-wave two-pion distribution amplitudes have been fixed in
Refs.~\cite{prd91-094024,plb763-29}. Based these work, we have studied the $S$-wave resonance contributions
to the decays $B^0_{(s)}\to \eta_c\pi^+\pi^-$~\cite{epjc76-675}, $B^0_{s}\to \psi(2s) \pi^+\pi^-$~\cite{1701.02941},
and the  $P$-wave resonance ($\rho(770)$) contributions  to $B^0_{(s)}\to (D/P) \rho\to (D/P)
\pi \pi$ decays~\cite{1611.08786,prd95-056008} with $D$ represents the charmed $D$ mesons and the $P$ stands for the light
pseudoscalar mesons: $\pi, K, \eta$ or $\eta^\prime$.

Up to now, several decay modes of the $B$ and $B_s$ mesons to the charmonium state plus pion pair,
like $B^0\to J/\psi \pi^+\pi^-$~\cite{prl90-091801,prd87-052001,prd90-012003,plb742-38}, $B_s^0\to J/\psi \pi^+\pi^-$~
\cite{prd86-052006,prd89-092006}, $B_{(s)}^0\to \psi(2S) \pi^+\pi^-$\cite{npb-871403} and
$B_s^0\to \eta_c \pi^+\pi^-$~\cite{1702.08048}, have been measured by BaBar and LHCb Collaboration.
With the continuous running of the LHCb experiment, more data of such $B/B_s$ decays with the inclusion
of various excited charmonium states ( $\eta_c(2S)$ etc.) will be collected. It is therefore interesting to
study such decay modes theoretically.
In this work, we will study the $S$-wave resonance contributions to $B^0_{(s)}\to \eta_c{(2S)}f_0(X)\to \eta_c(2S)\pi^+\pi^-$
decays and give our predictions for the branching fractions of the considered decay modes.

This paper is organized as follows. In Sec.~II, we give a brief introduction for the theoretical framework.
The numerical values, some discussions and the conclusions will be given in last two sections.

\section{The theoretical framework}\label{sec:2}

In the $B^0_{(s)}\to \eta_c{(2S)}\pi^+\pi^-$ decays, by using of the
light-cone coordinates and in the rest frame of $B^0_{(s)}$ meson,
the momentum of $B^0_{(s)}$, the pion pair and $\eta_c{(2S)}$ could be chosen as
\begin{eqnarray}\label{mom-pBpp3}
p_{B}=\frac{m_{B}}{\sqrt2}(1,1,0_{\rm T}),~\quad p=p_1+p_2=\frac{m_{B}}{\sqrt2}(1-r^2,\eta,0_{\rm T}),~\quad
p_3=\frac{m_{B}}{\sqrt2}(r^2,1-\eta,0_{\rm T})
\end{eqnarray}
where $\eta=\omega^2/[(1-r^2)m^2_{B}]$,
$r=m_{\eta_c{(2S)}}/m_{B}$ and $\omega^2=p^2$ means the squared invariant mass of the pion pair.
The momenta for the spectators in the ${B^0_{(s)}}$ meson, the pion pair, and the ${\eta_c{(2S)}}$ meson read as
\begin{eqnarray}\label{mom-B-k}
k_{B}=\left(0,\frac{m_{B}}{\sqrt2}x_{B},k_{BT}\right),~~
k=\left(\frac{m_{B}}{\sqrt2}z(1-r^2),0,k_{\rm T}\right),~~
k_3=\left(\frac{m_{B}}{\sqrt2}r^2x_3,\frac{m_{B}}{\sqrt2}(1-\eta)x_3,k_{3{\rm T}}\right),
\end{eqnarray}
where the momentum fractions $x_{B}$, $z$, and $x_3$ run from zero to unity.

\begin{figure}[]
\begin{center}
\vspace{-2cm} \centerline{\epsfxsize=17cm \epsffile{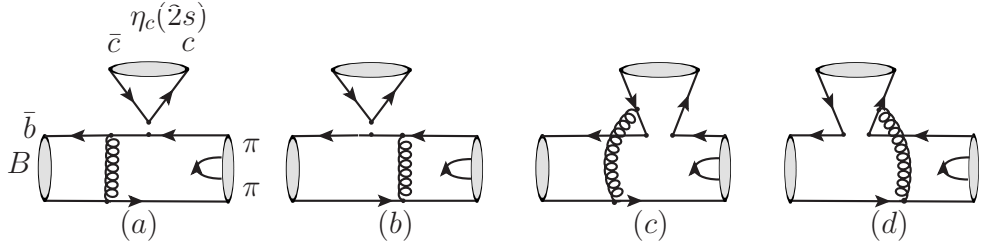}}
\vspace{-19cm} \caption{Typical Feynman diagrams contributing to the three-body decays
$B^0_s\to \eta_c{(2S)}\pi^+\pi^-$.} \label{fig:fig1}
\end{center}
\vspace{-1cm}
\end{figure}

The $S$-wave two-pion distribution amplitudes can be written as~\cite{prd91-094024,plb730-336}
\begin{eqnarray}
\Phi_{\pi\pi}^{S-wave}=\frac{1}{\sqrt{2N_c}}\left[p\!\!\!\slash\Phi_{v\nu=-}^{I=0}(z,\zeta,\omega^2)
+\omega\Phi_{s}^{I=0}(z,\zeta,\omega^2)+\omega( n \hspace{-2.2truemm}/_+ n \hspace{-2.2truemm}/_--1)
\Phi_{t\nu=+}^{I=0}(z,\zeta,\omega^2) \right],
\end{eqnarray}
with $n_+=(1,0,{\bf 0}_T)$, $n_-=(0,1,{\bf 0}_T)$ and the $\pi^+$ meson momentum
fraction $\zeta=p^+_1/p^+$. Their asymptotic forms are parameterized as~\cite{prd91-094024}
\begin{eqnarray}
\Phi_{v\nu=-}^{I=0}&=&\frac{9F_{s}(\omega^2)}{\sqrt{2N_c}}a_2^{I=0}z(1-z)(1-2z),~~
\Phi_{s}^{I=0}=\frac{F_{s}(\omega^2)}{2\sqrt{2N_c}},~~
\Phi_{t\nu=+}^{I=0} =\frac{F_{s}(\omega^2)}{2\sqrt{2N_c}}(1-2z),
\end{eqnarray}
with the time-like scalar form factor $F_{s}(w^2)$ and the Gegenbauer coefficient $a_2^{I=0}=0.2\pm0.2$.

The expressions of the time-like scalar form factor $F_{s}(\omega^2)$ associated with the $s\bar s$ component
of both $f_0(980)$ and $f_0(1500)$, and $d\bar d$ component of $f_0(500)$ can be found in Ref.~\cite{prd91-094024}.
Following the LHCb collaboration~\cite{prd86-052006,prd89-092006,prd87-052001,prd90-012003},
the Breit$-$Wigner (BW) formula for the $f_0(500)$ and $f_0(1500)$  resonances will be used to parameterize the time-like
scalar form factors in the two-pion distribution amplitudes, which include both
the resonant and non-resonant contributions of the $\pi\pi$ pair.
For $f_0(980)$, however, the Flatt\'e model \cite{plb63-228} will be used since $f_0(980)$ is close to the
$K\bar K$ threshold and the BW formula does not work well for this meson ~\cite{jpg34-151,plb63-228}.
We know that there exist some disputations about the nature of the meson $f_0(500)$ due to its wide
shape. Following the same treatment of $f_0(500)$ as LHCb collaboration~\cite{prd92-032002}, we here
also parameterize its contribution to the scalar form factor in the Bugg resonant line-shape~\cite{jpg34-151}
\begin{equation}
R_{f_0(500)}(s) = m_r \Gamma_1 (s)\; \left[m_r^2 - s - g_1^2\frac{s-s_A}{m_r^2 - s_A}
\left [ j_1(s) - j_1(m_r^2) \right ]
-i m_r \sum_{i=1}^{4}\Gamma_i (s) \right]^{-1} ,
\end{equation}
with the following relevant parameters
\begin{eqnarray}
 m_r\Gamma_1(s) &=& g_1^2 \frac{s-s_A}{m_r^2 - s_A}\rho_1(s),  \nonumber\\
g_1^2(s) & =& m_r(b_1 + b_2s)\exp(-(s-m_r^2)/A), \nonumber\\
j_1(s) & =& \frac{1}{\pi} \left[2+ \rho_1 \textrm{ln} \left( \frac{1-\rho_1}{1+\rho_1}\right) \right], \nonumber\\
m_r\Gamma_2(s) &=& 0.6g_1^2(s)(s/m_r^2)\exp(-\alpha|s-4m_K^2|)\rho_2(s), \nonumber\\
m_r\Gamma_3(s) &=& 0.2g_1^2(s)(s/m_r^2)\exp(-\alpha|s-4m_{\eta}^2|)\rho_3(s), \nonumber\\
m_r\Gamma_4(s) &=& m_rg_{4\pi}\rho_{4\pi}(s)/\rho_{4\pi}(m_r^2), \nonumber\\
 \rho_{4\pi}(s)  &=& 1/\left[1+\exp(7.082 - 2.845s)\right].
\end{eqnarray}
In the numerical calculation, we set $m_r$ = $0.953~GeV$, $s_A = 0.41 \ m_{\pi}^2$,
$b_1 = 1.302~GeV$, $b_2 = 0.340~GeV^{-1}$, $A = 2.426~GeV^2$ and $g_{4\pi} = 0.011~GeV$~\cite{jpg34-151}.
The phase-space factors of the decay channels $\pi\pi$, $KK$ and
$\eta\eta$ are defined as $\rho_i(s) = \sqrt{1 - 4m^2_i/s}$ with $i=1,2,3$ for
$\pi,K$ and $\eta$ respectively. It is worth of mentioning that another description of pion-pion form
factors were introduced in Ref.~\cite{prd83-074004,prd89-053015}.

For the $B^0_{(s)}$ mesons, we use the same distribution amplitudes  $\phi_B(x, b)$ in the $b$
space as being used for example in Ref.~\cite{epjc76-675},
\begin{eqnarray}
\Phi_B= \frac{i}{\sqrt{2N_c}} ({ p \hspace{-2.0truemm}/ }_B +m_B) \gamma_5 \phi_B ({\bf k_1}) \;. \label{bmeson}
\end{eqnarray}
The distribution amplitude is chosen as
\begin{eqnarray}
\phi_B(x,b)&=& N_B x^2(1-x)^2\mathrm{exp} \left  [ -\frac{M_B^2\ x^2}{2 \omega_{B}^2} -\frac{1}{2} (\omega_{B}\; b)^2\right] \;.
 \label{phib}
\end{eqnarray}
In the numerical calculation,
we also use the shape parameter $\omega_B = 0.40 \pm 0.04$~GeV with $f_B=0.19$~GeV
for $B^0$ decays, and $\omega_{B_s} = 0.50 \pm 0.05$~GeV with $f_{B_s}=0.236$~GeV
for $B^0_s$ decays ~\cite{epjc76-675}.

As the first radial excitation of the $\eta_c$ charmonium ground state, $\eta_c(2S)$ is observed firstly by
the Belle collaboration in $B$ decays~\cite{prl89-102001,prl92-142002}. The harmonic-oscillator wave function with the principal
quantum number $n=2$ and the orbital angular momentum $l=0$ is defined as~\cite{prd71-114008}
\begin{eqnarray}\label{eq:wavve}
\langle \eta_c(2S)|\bar{c}(z)_{\alpha}c(0)_{\beta}|0\rangle =
-\frac{i}{\sqrt{2N_c}}\int_0^1 dxe^{ixp_3\cdot z}
\left [(\gamma_5\rlap{/}{p_3})_{\alpha\beta}\psi^v(x,b)+m(\gamma_5)_{\alpha\beta}\psi^s(x,b)\right ].
\end{eqnarray}
The asymptotic models for the twist-2 distribution amplitudes $\psi^{v}$, and the
twist-3 distribution amplitudes $\psi^{s}$  for the radially excited $\eta_c(2S)$
is parameterized as~\cite{epjc75-293}
\begin{eqnarray}\label{eq:wave}
\Psi^v(x,b)&=&\frac{f_{\eta_c(2S)}}{2\sqrt{2N_c}}N^{v} x\bar{x}\mathcal {T}(x)
e^{-x\bar{x}\frac{m_c}{w}[w^2b^2+(\frac{x-\bar{x}}{2x\bar{x}})^2]},\nonumber\\
\Psi^s(x,b)&=&\frac{f_{\eta_c(2S)}}{2\sqrt{2N_c}}N^s \mathcal {T}(x)
e^{-x\bar{x}\frac{m_c}{w}[w^2b^2+(\frac{x-\bar{x}}{2x\bar{x}})^2]},
\end{eqnarray}
with the function $\mathcal {T}(x)=1-4b^2m_cw x\bar{x}+ m_c(x-\bar{x})^2/(w x\bar{x} )$ and the same normalization
conditions as the $B^0_{(s)}$ mesons: $ \int_0^1\Psi^{i}(x,b=0)d x = f_{\eta_c(2S)} /(2\sqrt{6})$.
And we also choose $f_{\eta_c(2S)}=0.243^{+0.079}_{-0.111}$ GeV and $w=0.2\pm0.1$ GeV as in Ref.~\cite{epjc75-293}.


In the PQCD factorization approach, there are four kinds of emission
Feynman diagrams for the $B^0_{(s)}\to \eta_c(2S)\pi^+\pi^-$ as illustrated in FIG.~1,
where (a) and (b) are factorizable diagrams, while (c) and (d) are the non-factorizable
ones.
We will use $F^{LL}, F^{LR}, F^{SP}$ and $M^{LL},M^{LR},M^{SP}$ to
describe the contributions of the factorizable [FIG. 1(a) and 1(b)] and
non-factorizable [FIG. 1(c) and 1(d)] emission
diagrams with the $(V-A)(V-A), (V-A)(V+A)$,
and $(S-P)(S+P)$ currents, respectively.
The total decay amplitudes for the considered decays can therefore be
written as
\begin{eqnarray}
\mathcal{A}({B^0_{(s)}\to \eta_c(2S)\pi^+\pi^-})&=&V^*_{cb}V_{cd(cs)}
\big[(C_1+\frac{C_2}{3})F^{LL} +C_2M^{LL}\big]-V^*_{tb}V_{td(ts)}\big[(  C_3+\frac{C_4}{3}+C_9+\frac{C_{10}}{3} )F^{LL}\nonumber\\
&+&(C_5+\frac{C_6}{3}+C_7+\frac{C_8}{3}\ )F^{LR}+ ( C_4+C_{10} )M^{LL}+(C_6+C_8)M^{SP} \big ],
\end{eqnarray}
where $C_i(\mu)(i=1, . . . ,10)$ are Wilson coefficients at the
renormalization scale $\mu$. For simplicity, we denote the distribution
amplitudes $\Phi_{v\nu=-}^{I=0}(z,\zeta,\omega^2)$
$[\Phi_{s}^{I=0}(z,\zeta,\omega^2), \Phi_{t\nu=+}^{I=0}(z,\zeta,\omega^2)]$
by $\phi_0$ $(\phi_s, \phi_\sigma)$ below.
From Fig.~1(a) and 1(b), we find
\begin{eqnarray}
 F^{LL}&=&8\pi C_F m^4_B f_{\eta_c(2S)}\int_0^1 dx_B dz
\int_0^\infty b_B db_B b db \phi_B(x_B,b_B)\nonumber\\
&\times&\bigg\{\bigg[\sqrt{\eta(1-r^2)}\Big[ \left ( (1-2z)(1-\eta)+r^2(1+2z(1-\eta)) \right )
(\phi_s-\phi_\sigma)+2(1-\eta)( 1-2(1-r^2)z )\phi_\sigma\Big]\nonumber\\
&+&\left[(1+z)(1-\eta)-r^2(1-2\eta+2z(1-\eta))\right]\phi_0\bigg]E_e(t_a)h_a(x_B,z,b_B,b)+\bigg[2\sqrt{\eta(1-r^2)}\nonumber\\
&\times&\left[1-\eta-r^2(1+x_B-2\eta)\right]\phi_s
+(1-r^2)\left[r^2(x_B-\eta)-(1-\eta)\eta \right]\phi_0\bigg]E_e(t_b)h_b(x_B,z,b_B,b)
\bigg\},\\
F^{LR}&=&-F^{LL},
\label{exp-F-LL}
\end{eqnarray}
with a color factor $C_F=4/3$.
The explicit expressions of the hard functions $h_a$ and $h_b$, the evolution
factors $E_e(t_i)$ including the Sudakov exponents and
the hard scales $(t_a, t_b)$ can be found for example in Ref.~\cite{prd91-094024}.
Following the same procedure, one can obtain the explicit expressions for decay amplitude
$M^{LL}$, $M^{LR}$ and $ M^{SP}$ from the evaluation of Fig.~1(c) and 1(d).


\section{Numerical results}\label{sec:3}
In numerical calculations, the following input parameters are
used implicitly. The QCD scale, masses and decay constants are in unit of GeV ~\cite{PDG2016}:
\begin{eqnarray}
\Lambda^{(f=4)}_{\overline{MS} } &=&0.25, \quad m_{B^0_s} = 5.367, \quad m_{B^0} = 5.280, \quad  M_{_{\eta_c(2S)}}= 3.639;\nonumber\\
m_\pi^{\pm}&=&0.140, \quad m_\pi^0=0.135, \quad m_c=1.27,
 \quad\tau_{B^0} = 1.520~{\; ps}, \quad\tau_{B^0_s} = 1.510~{\; ps}.
\label{eq:mass}
\end{eqnarray}
The Wolfenstein parameters for the CKM matrix elements read as~\cite{PDG2016}
\begin{eqnarray}
\lambda = 0.22506\pm 0.00050, \ \ A= 0.811 \pm 0.026 \ \ \bar{\rho} = 0.124_{-0.018}^{+0.019},
\ \ \bar{\eta}= 0.356\pm 0.011. \ \
\end{eqnarray}

The differential branching ratio for the $B^0_{(s)}\to \eta_c{(2S)}\pi^+\pi^-$ decay can be written as~\cite{prd91-094024}
\begin{eqnarray}
\frac{d{\cal B}}{d\omega}=\tau_{B}\frac{\omega|\overrightarrow{p_1}|
|\overrightarrow{p_3}|}{4(2\pi)^3m^3_{B}}|{\cal A}|^2,
\label{expr-br}
\end{eqnarray}
with the $B^0_{(s)}$ meson mean lifetime $\tau_{B}$.
The kinematic variables $|\overrightarrow{p_1}|$
and $|\overrightarrow{p_3}|$ denote the magnitudes of the $\pi^+$ and $\eta_c{(2S)}$
momenta in the center-of-mass frame of the pion pair,
\begin{eqnarray}
 |\overrightarrow{p_1}|=\frac12\sqrt{\omega^2-4m^2_{\pi^\pm}},~~~~
 |\overrightarrow{p_3}|=\frac{1}{2\omega}
  \sqrt{\left[m^2_{B}-(\omega+m_{\eta_c{(2S)}})^2 \right]\left[m^2_{B}-(\omega-m_{\eta_c{(2S)}})^2 \right]}.
\end{eqnarray}

\begin{figure}[thb]
\begin{center}
\vspace{-0.5cm}
\centerline{\epsfxsize=9.5cm \epsffile{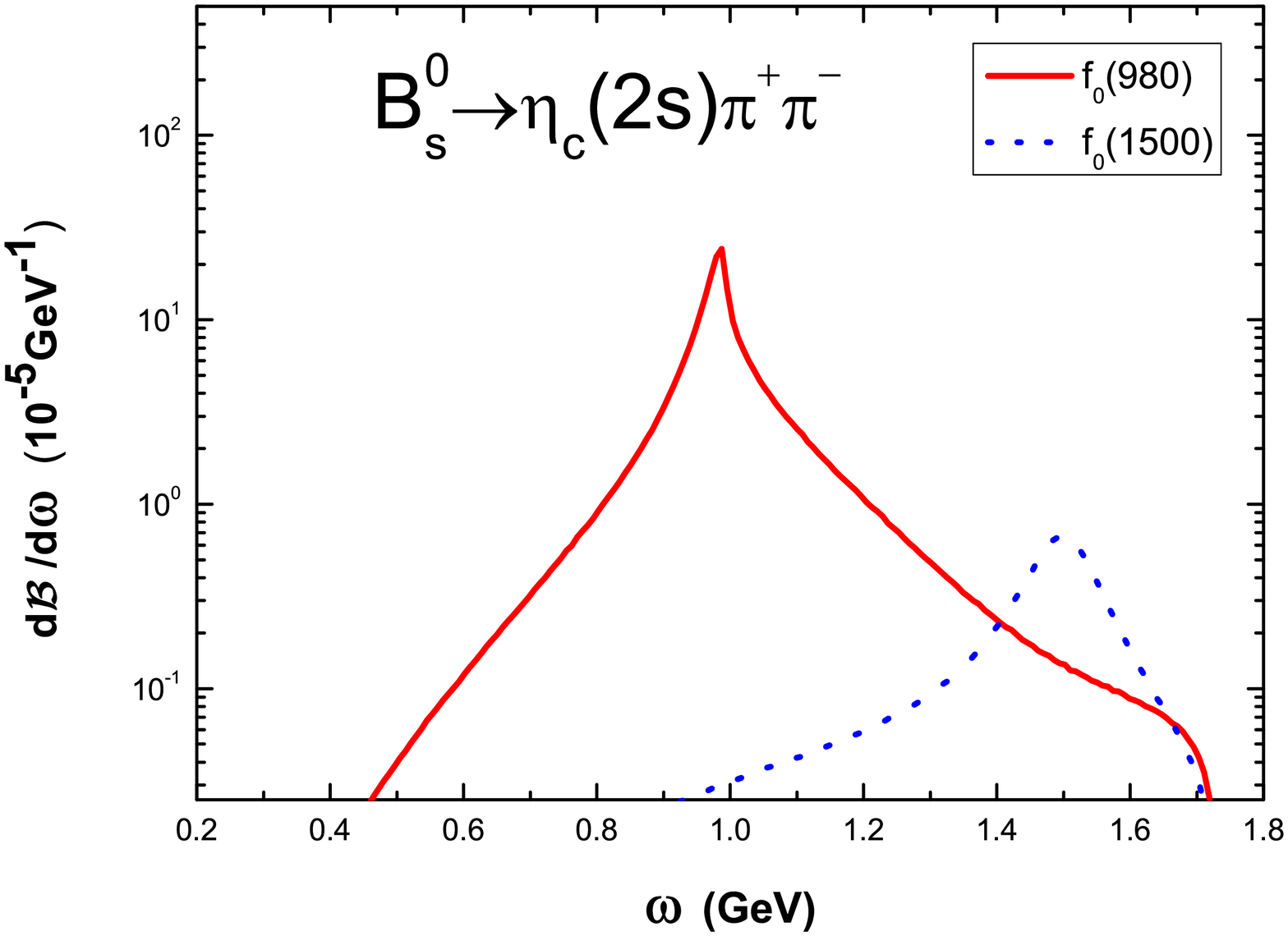}
\epsfxsize=9.5cm \epsffile{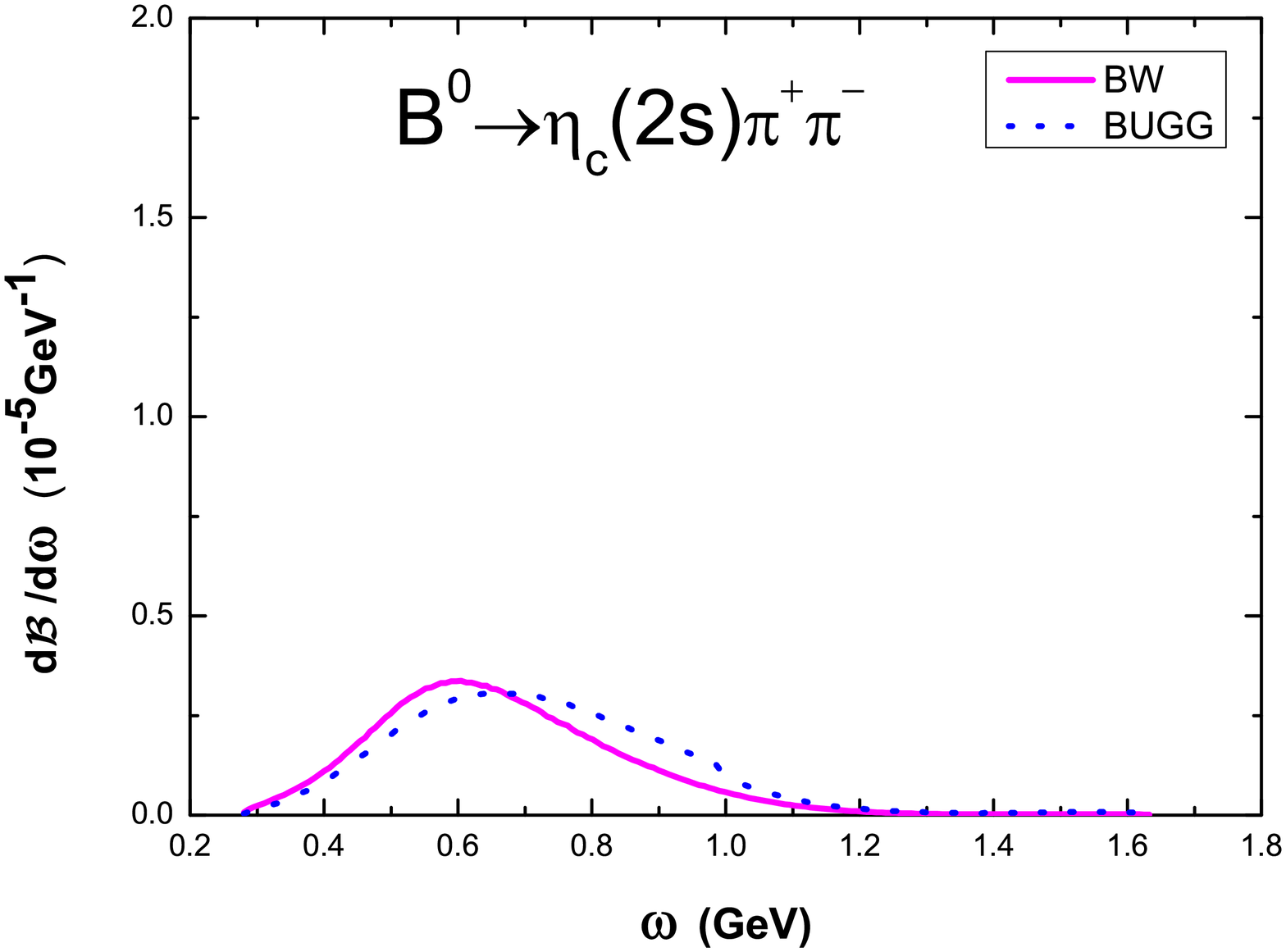}  }
(a)\hspace{9cm}(b)
\caption{ The $\omega$-dependence  of $d{\cal B}/d\omega$
for (a) the contribution from resonance $f_0(980)$ and $f_0(1500)$ for $B^0_s\to {\eta_c(2S)}\pi^+\pi^-$ decay;
and (b) the contribution from $f_0(500)$ for $B^0\to {\eta_c(2S)}\pi^+\pi^-$ decay.}
\label{figs-dep}
\end{center}
\end{figure}

From our numerical calculations, we find the following results:
\begin{itemize}
\item
In Fig.~2(a), we show the differential branching ratios $d{\cal B}/d\omega$ for
$B^0_s\to {\eta_c(2S)}\pi^+\pi^-$ decay, where the solid curve and the dots curve shows  the contribution
from $f_0(980)$  and $f_0(1500)$ is taken into account, respectively.
In Fig.~2(b), we show the $\omega$-dependence of the differential decay rate $d{\cal B}/d\omega$
when the BW model (solid curve) and the Bugg's model (dots curve) are employed.
The allowed region of $\omega$ is $4m_\pi^2 \leq \omega^2 \leq (M_{B}-m_{\eta_c(2S)})^2$.

\item
For the decays $ B^0_s\to \eta_c{(2S)} f_0(X)\to \eta_c(2S) \pi^+\pi^-$,
when the contribution from $f_0(980)$ and $f_0(1500)$ are included respectively,
the PQCD predictions for the branching ratios $ {\cal B}(B^0_s\to \eta_c{(2S)} f_0(X)\to \eta_c(2S) \pi^+\pi^-)$
are of the form of
{\footnotesize
\begin{eqnarray}
{\cal B}(B^0_s\to \eta_c{(2S)} f_0(980)[f_0(980)\to\pi^+\pi^-])&=&\left(2.19
^{+0.69}_{-0.55}(\omega_{B^0_s})^{+0.50}_{-0.42}(a_2)^{+1.05}_{-0.45}(w)^{+0.36}_{-0.26}(f_{\eta_c(2S)})\right)\times10^{-5},\nonumber\\
{\cal B}(B^0_s\to \eta_c{(2S)} f_0(1500)[f_0(1500)\to\pi^+\pi^-])&=&\left(1.31
^{+0.08}_{-0.12}(\omega_{B^0_s})^{+0.39}_{-0.31}(a_2)^{+0.62}_{-0.56}(w)^{+0.77}_{-0.50}(f_{\eta_c(2S)})\right)\times10^{-6},
\label{pqcd-prediction}
\end{eqnarray} }
where the first two errors come from the uncertainty $\omega_{B_s} = 0.50 \pm 0.05$~GeV and
$a^{I=0}_2=0.2\pm 0.2$, the last two errors are from $w=0.2\pm0.1$ GeV and $f_{\eta_c(2S)}=0.243^{+0.079}_{-0.111}$ GeV
( the parameters in the wave function of $\eta_c(2S)$).
The errors from the uncertainties of other input parameters, for instance the CKM matrix elements,
are very small and have been neglected.

By taking into account the S-wave contributions from $f_0(980)$ and $f_0(1500)$ simultaneously,
we find the PQCD prediction for the total branching ratio:
\begin{eqnarray}
{\cal B}(B^0_s\to \eta_c{(2S)} (\pi^+\pi^-)_S)&=&\left(2.67
^{+0.74}_{-0.62}(\omega_{B^0_s})^{+0.61}_{-0.54}(a_2)^{+1.43}_{-0.60}(w)^{+0.47}_{-0.36}(f_{\eta_c(2S)})\right)\times10^{-5}.
\end{eqnarray}
It is easy to see that the dominant contribution comes from the resonance $f_0(980)$ $(82.0\%)$,
while the constructive interference between $f_0(980)$ and $f_0(1500)$  provide $\sim 13\%$ enhancement to the
total decay rate.  One can read out this information from Fig. 2(a) approximately.
When compared with the previous study for $B^0_s\to \eta_c (\pi^+\pi^-)_s$ in Ref.~\cite{epjc76-675},
we find that ${\cal B}(B \to \eta_c(2S) [\pi^+\pi^-]_s):{\cal B}(  B \to \eta_c [\pi^+\pi^-]_s)\approx 1:2$.

\item
For $B^0\to \eta_c{(2S)} f_0(500)\to  \eta_c{(2S)} \pi^+\pi^- $ decay,
the PQCD predictions based on the BW model or the Bugg's model for the parametrization of the wide $f_0(500)$
are the following:
\begin{eqnarray}
{\cal B}(B^0\to \eta_c{(2S)} f_0(500)[f_0(500)\to \pi^+\pi^-])_{\rm (BW)}&=& 1.40^{+0.92}_{-0.56}\times10^{-6} \\
{\cal B}(B^0\to \eta_c{(2S)} f_0(500)[f_0(500)\to \pi^+\pi^-])_{\rm (Bugg)}&=&1.53^{+0.97}_{-0.61}\times10^{-6},
\end{eqnarray}
where the major errors have been added in quadrature. One can see easily that the PQCD predictions obtained by
employing the BW model or the Bugg's model are very similar, the difference is only about $10\%$.

\item
Based on our previous studies of the quasi-two-body B meson decays involving $\rho$ meson \cite{plb763-29},
we get to know that the main contribution lies indeed  in the region around the pole mass of the $\rho$ resonance.
Because $\Gamma_{\eta_c(2S)}\approx 11.3$ MeV is much narrow than $\Gamma_\rho \approx 149$ MeV, it is reasonable
for us to assume that the possible effect due to the narrow width of $\eta_c(2S)$ is very small
and can be neglected safely.
\end{itemize}

\section{Summary}\label{sec:4}

In summary, we studied the quasi-two-body $B^0_{(s)}\to \eta_c{(2S)}(\pi^+\pi^-)_S$ decays in
the PQCD factorization approach by introducing the $S$-wave two-pion distribution amplitudes.
For $B^0_s\to \eta_c{(2S)}f_0(X) \to \eta_c{(2S)} \pi^+\pi^- $ decay, the contributions from the $S$-wave resonance $f_0(980)$
and $f_0(1500)$ were taken into account, but the $f_0(980)$ provide the dominant contribution to the PQCD prediction:
${\cal B}(B^0_s\to \eta_c{(2S)} (\pi^+\pi^-)_S)=\left( 2.67^{+1.78}_{-1.08} \right)\times10^{-5}$.
For $B^0\to \eta_c{(2S)} f_0(X) \to \eta_c{(2S)} \pi^+\pi^- $ decay, the contribution from $f_0(500)$
was taken into account, the PQCD prediction for its decay rate is
$ \left( 1.40^{+0.92}_{-0.56} \right)\times10^{-6}$ in the BW model or
$ \left( 1.53^{+0.97}_{-0.61} \right)\times10^{-6}$ in the Bugg's model.
These PQCD predictions for the branching ratios of the considered decays can be measured and
tested at the near future LHCb and/or Belle-II experiments.

\begin{acknowledgments}

Many thanks to Hsiang-nan Li, Cai-Dian L\"u, Xin Liu, Rui Zhou and Wei Wang for valuable discussions.
This work was supported by the National Natural Science Foundation of China under the No.~11235005 and No.~11547038.

\end{acknowledgments}


\end{CJK*}
\end{document}